\documentclass[preprint,aps]{revtex4}

\newcommand {\bc}{\begin{center}}
\newcommand {\ec}{\end{center}}
\newcommand {\bea}{\begin{eqnarray}}
\newcommand {\eea}{\end{eqnarray}}
\newcommand {\be}{\begin{equation}}
\newcommand {\ee}{\end{equation}}

\def\lsim{\mathrel{\rlap{\lower4pt\hbox{\hskip1pt$\sim$}}
    \raise1pt\hbox{$<$}}}               
\def\gsim{\mathrel{\rlap{\lower4pt\hbox{\hskip1pt$\sim$}}
    \raise1pt\hbox{$>$}}}                
\usepackage{graphicx}

\begin{document}


\title{Dissipative fluid dynamics for the dilute Fermi gas at unitarity:
Free expansion and rotation}

\author{T.~Sch\"afer}

\affiliation{Department of Physics, North Carolina State University,
Raleigh, NC 27695}

\begin{abstract}
 We investigate the expansion dynamics of a dilute Fermi gas
at unitarity in the context of dissipative fluid dynamics. Our 
aim is to quantify the effects of shear viscosity on the time 
evolution of the system. We compare exact numerical solutions
of the equations of viscous hydrodynamics to various approximations
that have been proposed in the literature. Our main findings
are: i) Shear viscosity leads to characteristic features in the 
expansion dynamics; ii) a quantitative description of these
effects has to include reheating; iii) dissipative effects are
not sensitive to the equation of state $P(n,T)$ as long as 
the universal relation $P=\frac{2}{3}{\cal E}$ is satisfied; 
iv) the expansion dynamics mainly constrains the cloud average 
of the shear viscosity. 

\end{abstract}

\maketitle

\section{Introduction}
\label{sec_intro}

 A cold, dilute Fermi gas in which the scattering length can 
be tuned to infinity by means of a Feshbach resonance provides
a new realization of a strongly correlated quantum fluid
\cite{Bloch:2007,Giorgini:2008,Chin:2009}. On resonance the two-body
scattering amplitude saturates the s-wave unitarity bound, 
and the corresponding many-body system is referred to as the 
Fermi gas at unitarity. An important manifestation of strong 
correlations is the observation of nearly ideal hydrodynamic 
flow \cite{OHara:2002}. In the Fermi gas at unitarity nearly 
ideal flow was observed in samples containing as few as $10^5$ 
atoms, with an interparticle spacing between the atoms on the 
order of several $10^3$ $\AA$, much larger than the range of 
the interaction.

 This result implies that dissipative effects must be very small 
\cite{Schafer:2009dj}. In a normal fluid dissipative phenomena 
are governed by three transport coefficients, the shear viscosity 
$\eta$, the bulk viscosity $\zeta$, and the thermal conductivity 
$\kappa$. The Fermi gas at unitarity is scale invariant and the 
bulk viscosity is zero \cite{Son:2005tj,Escobedo:2009bh}. Most of 
the experiments conducted so far, like collective oscillations and 
the expansion from a deformed trap, involve scaling flows. In these 
flows the cloud remains nearly isothermal and the experiment is not 
sensitive to thermal conductivity \cite{Schaefer:2009px,Braby:2010ec}. 
The observation of nearly ideal flow therefore requires that the 
shear viscosity is very small. 

 From a theoretical point of view we know that the shear viscosity 
of a weakly interacting gas scales as $\eta\sim \bar{p}/\sigma$, 
where $\bar{p}$ is the mean momentum and $\sigma$ is the scattering
cross section. This implies that the shear viscosity of a strongly 
interacting gas is expected to be small. At unitarity the cross 
section is $\sigma=4\pi/k^2$, where $k$ is the momentum transfer. 
The average cross section in a thermal gas is $\sigma\simeq
\frac{4\pi}{3} mT$, and the shear viscosity is $\eta\simeq 
\frac{15}{32\sqrt{\pi}}(mT)^{3/2}$ \cite{Bruun:2005,Bruun:2006}. 
This result is reliable as long as $T$ is much larger than the 
critical temperature $T_c$ for superfluidity. Below $T_c$ the 
nature of the excitations changes and $\eta\sim 1/T^5$ 
\cite{Rupak:2007vp}. These results indicate that the shear 
viscosity has a minimum at a temperature on the order of $T_c$, 
but kinetic theory cannot reliably predict how small the minimum value 
of the shear viscosity is. The region near $T_c$ can be studied
using sum rules \cite{Taylor:2010ju}, or with the help Kubo's formula
and many body perturbation theory \cite{Enss:2010qh}.

 It has been argued that the quantum mechanical uncertainty relation 
implies a lower bound $\eta/n\sim \bar{p}l_{\it mfp}\geq \hbar$ 
\cite{Danielewicz:1984ww}. Here, $n$ is the density, $l_{\it mfp}
\sim 1/(n\sigma)$ is the mean free path, and $\hbar$ is Planck's 
constant. A more precise bound has emerged from the study of 
holographic dualities in string theory. Kovtun, Son, and Starinets 
proposed that the shear viscosity to entropy density ratio $\eta/s$ 
is bounded from below by $\hbar/(4\pi k_B)$ \cite{Kovtun:2004de}, 
where $k_B$ is Boltzmann's constant.

 Simple estimates of the shear viscosity to entropy density ratio
based on experimental data indicate that $\eta/s$ in the unitary 
Fermi gas is indeed close to the proposed bound. The first attempt 
to determine $\eta/s$ from data was based on the damping of collective 
modes \cite{Gelman:2004,Schafer:2007pr,Turlapov:2007}. The damping 
constant can be related to the ratio $\dot{E}/E$, where $E$ is 
the total energy of the mode and $\dot{E}$ is the energy dissipated
by viscous effects. For a scaling flow $\dot{E}$ is proportional 
to the spatial integral of the shear viscosity. It was found that 
the ratio of the trap averages of $\eta$ and $s$ has a minimum 
value of $\langle \eta\rangle/\langle s\rangle \simeq 0.5$
\cite{Schafer:2007pr,Turlapov:2007}, where we have set $\hbar
=k_B=1$.

 More recent analyses are based on the dynamics of an expanding cloud 
\cite{Clancy:2007,Clancy:2008,Thomas:2009zz,Schaefer:2009px}. These
studies utilize approximate solutions of the equations of 
dissipative hydrodynamics. The first approximations is that entropy 
is assumed to be conserved, which is equivalent to the assumption
that the system is in contact with an energy sink that removes the 
heat generated by dissipative effects. The second approximation 
is that the Navier-Stokes equation is converted into a set of 
ordinary differential equations by taking moments. If only the 
lowest moments are included then the result is only sensitive
to the spatial integral of the shear viscosity. An analysis of 
the expansion of a rotating cloud gives values as small as $\langle 
\eta\rangle/\langle s\rangle \simeq 0.2$ \cite{Thomas:2009zz}.
A more refined approximation was used in \cite{Cao:2010} to 
analyze the high temperature limit of the shear viscosity. We
will describe this method in Sect.~\ref{sec_app}. 

 The goal of our present work is to test these approximations
by performing numerical studies of the equations of viscous 
hydrodynamics for an expanding cloud of a dilute Fermi gas 
at unitarity. This paper is structured as follows. In 
Sect.~\ref{sec_hydro} and \ref{sec_init} we introduce the 
equations of dissipative hydrodynamics for a scale invariant 
non-relativistic fluid. We discuss exact solutions for the 
ideal case and approximate solutions for the viscous case in 
Sects.~\ref{sec_exact} and \ref{sec_app}. Numerical solutions
are discussed in Sect.~\ref{sec_res} and our conclusions are 
summarized in Sect.~\ref{sec_sum}.

\section{Dissipative hydrodynamics}
\label{sec_hydro}

 We will consider the unitary Fermi gas in the normal phase. In 
that case there are five hydrodynamic variables, the mass density 
$\rho$, the flow velocity $\vec{v}$, and the energy density ${\cal E}$. 
These variables satisfy four hydrodynamic equations, the continuity 
equation, the Navier-Stokes equation, and the equation of 
energy conservation,   
\bea
\label{hydro1}
\frac{\partial \rho}{\partial t} 
   + \vec{\nabla}\cdot\left(\rho\vec{v}\right)  &=& 0 , \\
\label{hydro2}
 \frac{\partial (\rho v_i)}{\partial t}  
   + \nabla_j\Pi_{ij} &=& 0, \\
\label{hydro3}
 \frac{\partial {\cal E}}{\partial t} 
 \; +\; \vec{\nabla} \cdot\vec{\jmath}^{\;\epsilon} &=& 0 .  
\eea 
The total energy density is the sum of the internal energy
density and kinetic energy density, ${\cal E}={\cal E}_0+\frac{1}{2}
\rho v^2$. These equations close once we supply constitutive 
relations for the stress tensor $\Pi_{ij}$ and the energy current 
$\jmath_i^{\;\epsilon}$ as well as an equation of state. The 
unitary Fermi gas is scale invariant and the equation of state 
is $P=\frac{2}{3}{\cal E}_0$. The stress tensor is given by 
\be 
 \Pi_{ij} = \rho v_i v_j + P\delta_{ij}+ \delta \Pi_{ij}\, ,
\ee
where $\delta\Pi_{ij}$ is the dissipative part. The dissipative
contribution to the stress tensor is $\delta\Pi_{ij}=-\eta
\sigma_{ij}$ with
\be 
 \sigma_{ij} = \left(\nabla_i v_j +\nabla_j v_i 
  -\frac{2}{3}\delta_{ij}(\nabla_k v_k)\right)\, ,
\ee
where $\eta$ is the shear viscosity and we have used the fact 
that the bulk viscosity of the unitary Fermi gas is zero. The
energy current is 
\be
 \jmath_i^{\;\epsilon} = v_iw+\delta\jmath_i^{\;\epsilon}\, ,
\ee 
where $w={\cal E}_0+P$ is the enthalpy density. The dissipative 
energy current is 
\be 
\delta \jmath_i^{\;\epsilon} = \delta\Pi_{ij} v_j - 
\kappa \nabla_i T\, , 
\ee
where $T$ is the temperature and $\kappa$ is the thermal conductivity. 
We note that the temperature $T=T(n,P)$ is a function of the density 
$n=\rho/m$ and the pressure. In order to determine $T$ we need the 
equation of state in the form $P=P(n,T)$. Universality implies that 
$P(n,T)=m^{-1}n^{5/3}f_n(mT/n^{2/3})$ where $f_n(x)$ is a universal function 
that has to be determined experimentally or using quantum Monte Carlo 
methods. The situation simplifies in the high temperature limit where 
$P=nT$. We provide a parameterization of $P(n,T)$ at all temperatures 
in Appendix \ref{sec_eos}. Universality also restrict the dependence
of the shear viscosity and thermal conductivity on the density and
the temperature. We can write 
\be 
\label{alpha_n}
\eta(n,T) = \alpha_n\left(\frac{mT}{n^{2/3}}\right)\, n \, , 
\hspace{1cm}
\kappa(n,T) = \sigma_n\left(\frac{mT}{n^{2/3}}\right)\, \frac{n}{m} \, , 
\ee
where $\alpha_n(y)$ and $\sigma_n(y)$ are universal functions of
$y=mT/n^{2/3}$. The relative importance of thermal and momentum
diffusion can be characterized in terms of a dimensionless ratio
known an the Prandtl number, ${\it Pr}=c_p\eta/(\rho\kappa)$, where
$c_p$ is the specific heat at constant pressure. In the high 
temperature limit $c_p=\rho/m$ and ${\it Pr}=\alpha_n/\sigma_n$. 
Kinetic theory predicts that in this limit ${\it Pr}=2/3$
\cite{Braby:2010ec}.

\section{Initial conditions and choice of units}
\label{sec_init}

 We will consider the expansion of a dilute Fermi gas after
release from a harmonic trap. The density distribution in the initial
state satisfies the equation of hydrostatic equilibrium, $\vec{\nabla}P
=-n\vec{\nabla}V$, where $V(x)=\frac{1}{2}m\omega_i^2x_i^2$ is the 
trapping potential. If the gas is isothermal then this equation is 
solved by the local density approximation
\be 
\label{lda}
 n_0(x)=n(\mu(x),T)\, , \hspace{1cm}
 \mu(x)= \mu-V(x)\, , 
\ee
where $n(\mu,T)$ is the density in thermal equilibrium. The function
$n(\mu,T)$ can be determined from the equation of state as explained
in appendix \ref{sec_eos}. The simplest case is the high temperature
limit. In this limit the initial density is a Gaussian
\be
\label{n_0_Gauss}
n_0(x)= n_0\exp\left(-\sum_i\frac{x_i^2}{R_i^2}\right)\, ,
\ee
with $R_i^2=(2T)/(m\omega_i^2)$ and $n_0= N/(\pi^{3/2}R_xR_yR_z)$.  
The total number of particles is denoted by $N$. In the following we 
will use a dimensionless coordinate variable $\bar{x}_i=x_i/x_0$ where  
\be
\label{x_0_def}
 x_0^2 = \frac{2}{3m} 
   \left(\frac{3N}{\omega_x\omega_y\omega_z}\right)^{1/3}  \, . 
\ee
This variable can be used for any initial condition, not just 
the Gaussian initial condition given in equ.~(\ref{n_0_Gauss}).
We will focus on axially symmetric traps with $\omega_x=\omega_y=
\omega_\perp$ and $\omega_z=\lambda\omega_\perp$. In the high 
temperature limit the dimensionless density $\bar{n}=nx_0^3$ is 
given by 
\be 
\label{n_0_barx}
 \bar{n}_0(\bar{x}) = \bar{n}_0 \exp\left( -\frac{E_F}{E_0}
   \left( \bar{x}^2+\bar{y}^2+\lambda^2\bar{z}^2\right)\right)\, , 
\ee
where $E_0$ is the total (potential and internal) energy of the 
trapped gas and $E_F=(3N\lambda)^{1/3}N\omega_\perp$. The 
central density is $\bar{n}_0=\lambda N(E_F/E_0)^{3/2}/\pi^{3/2}$.

 We can write the equations of hydrodynamics in dimensionless
variables by introducing a scaled time variable $\bar{t}= 
\omega_\perp t$ as well as a scaled velocity, energy density, 
and pressure, 
\be 
\label{bar_v}
\bar{v}_i=\frac{v_i}{x_0\omega_\perp}, \hspace{1cm}
\bar{\cal E} = \frac{x_0}{m\omega_\perp^2}\,{\cal E}, \hspace{1cm}
\bar{P}= \frac{x_0}{m\omega_\perp^2}\, P \, . 
\ee
The scaled mass density is $\bar{\rho}=\rho x_0^3/m$. Using these 
variables the equations of fluid dynamics, equ.~(\ref{hydro1}-\ref{hydro3}), 
remain unchanged except for the change from dimensionful to dimensionless 
hydrodynamic variables. The dimensionless shear viscosity is $\bar{\eta} = 
x_0\eta/(m\omega_\perp)$. We can write $\bar{\eta}=\bar{\alpha}_n
\bar{n}$ with 
\be
\label{bar_alpha}
\bar{\alpha}_n = \frac{3}{2}\frac{1}{(3\lambda N)^{1/3}}\,\alpha_n\, , 
\ee
where $\alpha_n$ is the universal function introduced in 
equ.~(\ref{alpha_n}). Finally, we can introduce a dimensionless
temperature and chemical potential, $\bar{T}=T/(m\omega_\perp^2x_0^2)$
and $\bar{\mu}=\mu/(m\omega_\perp^2x_0^2)$.

\section{Exact solutions}
\label{sec_exact}

 In order to test the accuracy of the numerical hydrodynamics 
code we have studied a number of exactly solvable test cases. 
In ideal hydrodynamics there are exact scaling solutions for 
the expansion from rotating and non-rotating traps. Consider 
a density profile of the form 
\be 
\label{n_scale}
 n(x,t) =  \frac{1}{b_x(t)b_y(t)b_z(t)}
        F\left(\frac{x^2}{b_x^2(t)} +\frac{y^2}{b_y^2(t)}
        +\frac{\lambda^2 z^2}{b_z^2(t)}\right)\, , 
\ee
where $F(x)$ is an arbitrary function and the scale parameters
$b_i(t)$ satisfy the initial condition $b_i(0)=1$. This ansatz
satisfies the continuity equation with a velocity field given 
by $v_i(x,t)=\alpha_i(t)x_i$ with $\alpha_i=\dot{b}_i/b_i$.
The initial condition for the pressure can be determined by 
integrating the equation of hydrostatic equilibrium
\be
\label{hydro_st}
 P_0(x)=-\int n_0(x)\vec{\nabla}V(x)\cdot d\vec{x}\, .
\ee
In the limit $T\gg T_F$ the function $F(x)$ is a Gaussian and
the initial pressure is determined by the ideal gas equation of
state, $P=nT$. In the initial state the temperature is constant 
and the chemical potential is parabolic. In ideal hydrodynamics
the evolution of the system preserves these properties. The 
Gibbs-Duhem relation $dP=nd\mu +sdT$ implies that the force 
$(\vec\nabla P)/n=\vec\nabla\mu$ is exactly linear at all times. 

 The Euler equation is equivalent to three coupled 
ordinary differential equations for the scale parameters
$b_i$. We get \cite{Menotti:2002,Schaefer:2009px}
\be 
\label{b_i_ode}
\ddot b_i = \frac{\omega_i^2}{(b_xb_yb_z)^{2/3}}\frac{1}{b_i}\, ,
\ee
with the initial conditions $b_i(0)=1$ and $\dot{b}_i(0)=0$.
In the case of axial symmetry this set of equations reduces to 
two independent equations for $b_\perp=b_x=b_y$ and $b_z$. These 
differential equations have to be solved numerically. 

 This solution can be generalized to an initial velocity field
that corresponds to a rotating trap. The velocity field can be
chosen to be irrotational, $\vec{v}=\alpha\, \vec{\nabla}(xz)$, 
or rigidly rotating, $\vec{v}=\Omega\,\hat{y}\times \vec{x}$. Here 
we have chosen the direction of the angular momentum to be in the 
$y$-direction. In the rotating case the profile function in 
equ.~(\ref{n_scale}) has to be generalized to include an off-diagonal
$xz$-term. In total one has to solve for ten functions, the four scale
parameters $b_x,b_y,b_z$ and $b_{xz}$, the chemical potential at 
the center of the trap, and five functions characterizing the 
velocity field, $\alpha_x,\alpha_y,\alpha_z,\alpha$ and $\Omega$.
The equations of motion are given in \cite{Edwards:2002,Schaefer:2009px}.

 There are no known exact solutions for an expanding gas in 
the dissipative case. However, we can find scaling solutions
to the continuity and Navier-Stokes equation if the local shear 
viscosity is of the form $\eta=\eta_0 P/T$, where $\eta_0$ is a
constant. Note that at high temperature $P=nT$ and $\eta=\eta_0 n$. 
Also, for any scaling solution $T={\it const}$ and $\vec{\nabla}\eta 
= \eta_0(\vec{\nabla}P)/T$, which implies that both the ideal and 
the dissipative forces are proportional to the gradient of the 
pressure. The Navier-Stokes equation then leads to the coupled set of 
equations
\bea
\label{ns_mom_1} 
\ddot b_\perp  &=& \frac{\omega_\perp^2}
   {(b_\perp^2 b_z)^{2/3} b_\perp}
   -  \frac{2\beta\omega_\perp}{b_\perp}
      \left( \frac{\dot b_\perp}{b_\perp} 
                - \frac{\dot b_z}{b_z} \right)\,  \\
\label{ns_mom_2}
\ddot b_z  &=& \frac{\omega_z^2}
   {(b_\perp^2 b_z)^{2/3} b_z}
   +  \frac{4\beta\lambda\omega_z }{b_z}
      \left( \frac{\dot b_\perp}{b_\perp} 
                - \frac{\dot b_z}{b_z} \right) ,
\eea
where we have specialized the solution to the case of axial
symmetry and we have defined $\beta=\eta_0\omega_\perp/(3T_0)$,
where $T_0$ is the initial temperature. We can write the 
parameter $\beta$ as
\be 
\label{beta_def}
\beta = \frac{\langle\alpha_n\rangle}{(3N\lambda)^{1/3}}
        \frac{1}{(E_0/E_F)} 
      = \frac{2}{3}\frac{\langle \bar\alpha_n\rangle}{(E_0/E_F)}\, ,
\ee
where $\langle \alpha_n\rangle$ is the trap average of $\alpha_n$, 
\be  
\label{alpha_av}
 \langle \alpha_n\rangle = \frac{1}{N} 
   \int d^3x\, \alpha_n\!\left(\frac{mT}{n_0(x)^{2/3}}\right)n_0(x)\, ,
\ee
and we have used equ.~(\ref{bar_alpha}). The solution of 
equ.~(\ref{ns_mom_1},\ref{ns_mom_2}) does not conserve energy. 
Instead, it satisfies a modified energy equation
\be
\label{heat_sink}
 \frac{\partial {\cal E}}{\partial t} 
  + \vec{\nabla} \cdot\vec{\jmath}^{\;\epsilon} 
 = - \frac{\eta}{2}\left(\sigma_{ij}\right)^2\, ,
\ee
which contains a sink that removes the heat generated by dissipative 
effects. This means that the scaling solution conserves entropy even 
if the shear viscosity is not zero. The produced entropy is removed by 
the heat sink. We also note that equ.~(\ref{ns_mom_1},\ref{ns_mom_2}) 
can be generalized to the rotating case, see \cite{Schaefer:2009px}.

\section{Viscous hydrodynamics: Approximate solutions}
\label{sec_app}

 If dissipative effects are mostly governed by viscous forces,
and reheating is not important, then the scaling solution introduced
in the previous section is a useful approximation to the full 
hydrodynamic equations. We will show below that for an expanding 
system this is not the case. A more useful approximation was 
recently proposed in \cite{Cao:2010}. We will assume that the 
local shear viscosity is proportional to the density, $\eta = 
\alpha_n n$, where $\alpha_n$ is a constant. The basic idea 
is to focus on the force $f_i=(\nabla_i P)/n$ rather than the 
pressure itself. The Navier-Stokes equation is 
\be
\label{ns_force}
m\left(\frac{\partial}{\partial t} +\vec{v}\cdot\vec{\nabla}\right)v_i
  =f_i + \frac{\nabla_j (\eta\,\sigma_{ij})}{n}\, . 
\label{eq:force}
\ee
With the help of the Navier-Stokes equation the energy equation can 
be written as 
\be
\label{q_force}
 \left(\frac{\partial}{\partial t} + \mathbf{v}\cdot\nabla +
    \frac{2}{3}\left(\vec{\nabla}\cdot\vec{v}\right)\right)f_i 
   + (\nabla_i v_j)f_j 
   -\frac{5}{3}\left(\nabla_i\nabla_j v_j\right)\frac{P}{n} 
  =  -\frac{2}{3}\frac{\nabla_i\,\dot{q}}{n},
\ee
where $\dot{q}=\frac{\eta}{2}(\sigma_{ij})^2$ is the heating 
rate. The basic idea is to assume that even in the dissipative
case the velocity field and the force remain linear in the 
coordinates. If the velocity is linear and $\eta\sim n$ then 
all terms in equ.~(\ref{ns_force}) are linear in $x_i$. Also, 
equ.~(\ref{q_force}) is independent of the pressure and all
the remaining terms are linear in $x_i$. We write $f_i=a_i x_i$,
$v_i=\alpha_i x_i$ (no sum over $i$) and use the scaling ansatz 
(\ref{n_scale}) for the density. The continuity equation requires
$\alpha_i=\dot{b}_i/b_i$. The scale parameters $a_i$ and $b_i$ 
are determined by the coupled equations
\bea
\label{ns_for_1} 
\frac{\ddot b_\perp}{b_\perp}  &=&  a_\perp
   -  \frac{2\beta\omega_\perp}{b^2_\perp}
      \left( \frac{\dot b_\perp}{b_\perp} 
                - \frac{\dot b_x}{b_x} \right)\, ,  \\
\label{ns_for_2}
\frac{\ddot b_z}{b_z}  &=& a_z
   +  \frac{4\beta\lambda\omega_z}{b^2_z}
      \left( \frac{\dot b_\perp}{b_\perp} 
                - \frac{\dot b_z}{b_z} \right)\, ,\\
\label{ns_for_3}
\dot{a}_\perp  &=& 
 \mbox{}-\frac{2}{3}\,a_\perp
   \left(5\,\frac{\dot{b}_\perp}{b_\perp} + \frac{\dot{b}_z}{b_z}\right)
 + \frac{8\beta\omega_\perp^2}{3b_\perp}
  \left(\frac{\dot{b}_\perp}{b_\perp} - \frac{\dot{b}_z}{b_z}\right)^2
  \, , \\
\label{ns_for_4}
 \dot{a}_z  &=& 
 \mbox{}-\frac{2}{3}\,a_z
   \left(4\,\frac{\dot{b}_z}{b_z} + 2\, \frac{\dot{b}_\perp}{b_\perp}\right)
 + \frac{8\beta\lambda\omega_z}{3b_z^2}
  \left(\frac{\dot{b}_\perp}{b_\perp} - \frac{\dot{b}_z}{b_z}\right)^2
  \, , 
\eea
where $\beta$ is defined in equ.~(\ref{beta_def}). The initial 
conditions are $b_\perp(0)=b_z(0)=1$, $\dot{b}_\perp(0)=\dot{b}_z(0)=0$ 
as before, and $a_\perp(0)=\omega_\perp^2$, $a_z(0)=\omega_z^2$.
We note that the solutions of equ.~(\ref{ns_for_1}-\ref{ns_for_4})
provide an exact solution of the continuity, Navier-Stokes, and 
energy conservation equation. The solution is approximate in the 
sense that one cannot in general find a consistent expression for the 
pressure $P(\mu,T)$ such that $P,{\cal E},n$ are related by 
thermodynamic identities.

\begin{figure}[t]
\bc\includegraphics[width=10cm]{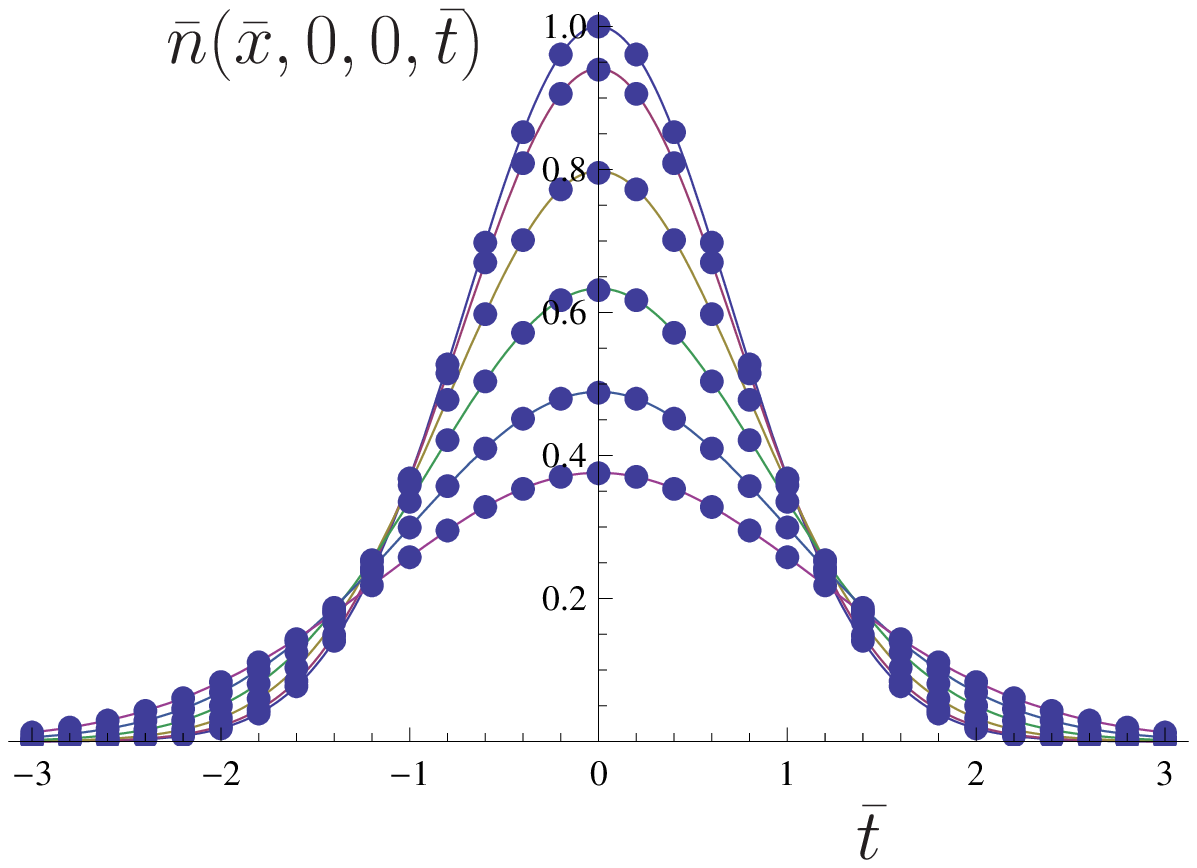}\ec
\bc\includegraphics[width=10cm]{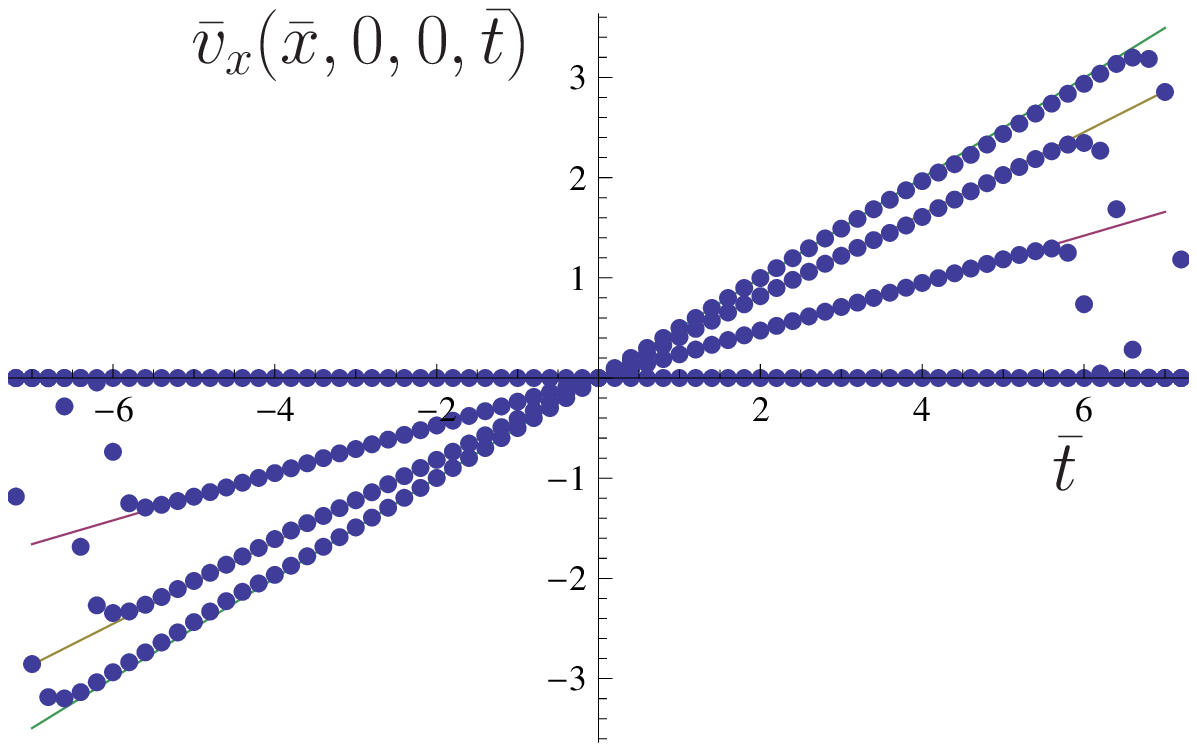}\ec
\caption{\label{fig_prof_tr}
Density and velocity profile of an expanding gas cloud in ideal 
hydrodynamics. The initial condition is a Gaussian density profile
with $E_0/E_F=1$. The top panel shows the density $\bar{n}(\bar{x},0,0,
\bar{t})$ for $\bar{t}=0.0,0.25,0.50,\ldots,1.25$. The solid lines 
show the analytic solution of the Euler equation, and the dots show 
numerical results. The bottom panel shows the velocity $\bar{v}_x
(\bar{x},0,0,\bar{t})$ for $\bar{t}=0.0,0.25,0.50,0.75$.}   
\end{figure}

\section{Viscous hydrodynamics: Numerical results}
\label{sec_res}
\subsection{Numerical tests}
\label{sec_test}

 Ideal hydrodynamic simulations were carried out using the VH1
code written by Blondin and Lufkin \cite{Blondin:1993}. VH1 uses
the PPMLR (Piecewise-Parabolic Method, Lagrangian-Remap) method 
developed by Colella and Woodward \cite{Woodward:1984,Colella:1984}.
The hydrodynamic equations are written in the form of conservation
laws and solved in Lagrangian coordinates. A Lagrangian time step
is followed by a piecewise parabolic remap onto an Eulerian grid.  
We have modified VH1 to include viscous corrections to the stress
tensor and the energy current. In the current work we have not 
included the effect of thermal conductivity. In the ideal case the 
cloud remains isothermal during the expansion and $\vec{\nabla}T=0$. 
Dissipative effects lead to non-zero temperature gradients, but if 
the viscosity and thermal conductivity are small then the corresponding 
correction to the energy current is second order in small quantities. 
We will verify this statement in Sect.~\ref{sec_reh}.

\begin{figure}[t]
\bc\includegraphics[width=10cm]{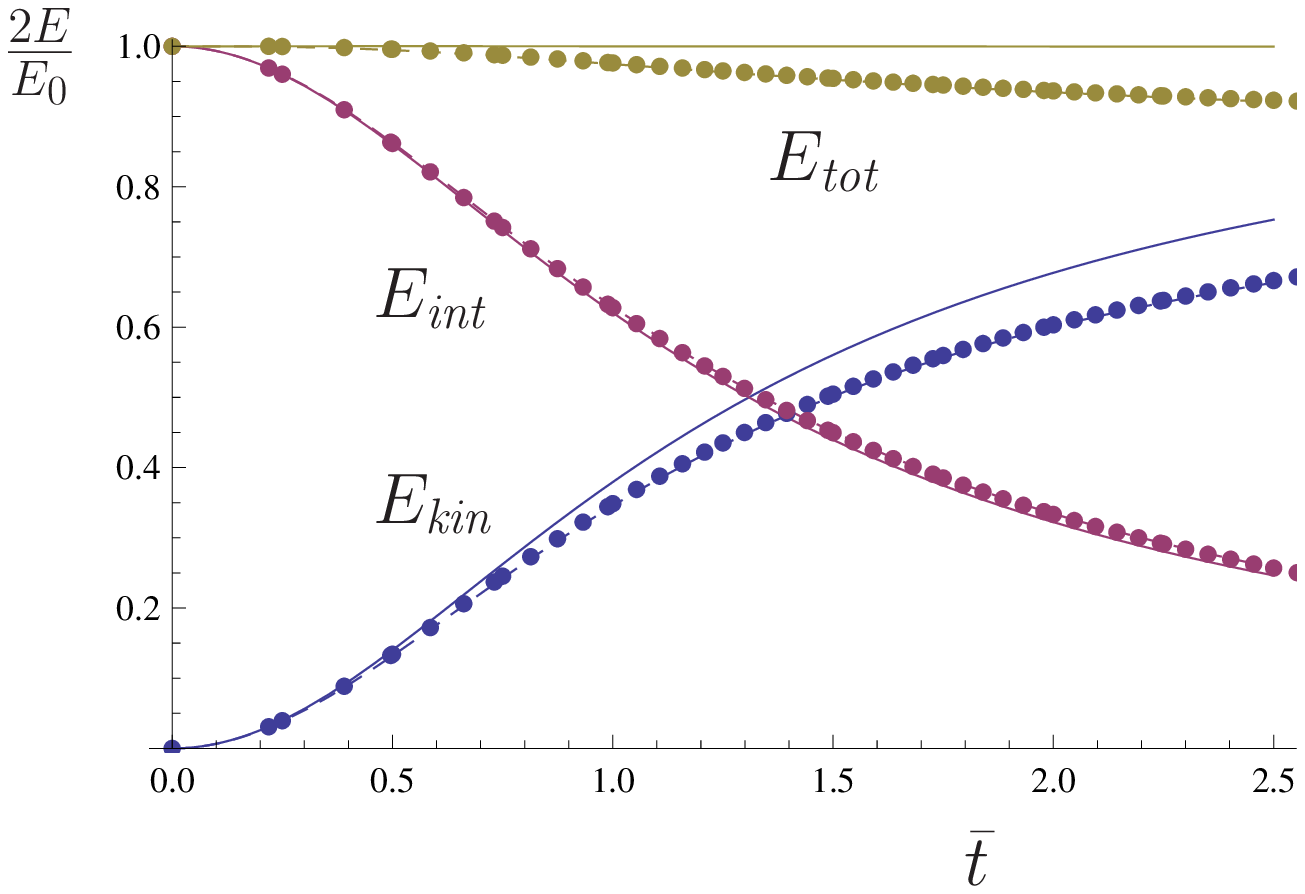}\ec
\caption{\label{fig_e_t_work}
Energy of the expanding gas cloud as a function of time. We show the
kinetic, internal, and total energy. The solid lines show the 
analytical result for the ideal evolution, the dashed lines show the
analytical result without reheating for $\beta=0.066$, and the points 
are from a numerical calculation with $E_0/E_F=1$ and $\bar\eta=
\bar\alpha_n \bar{n}$ with $\bar\alpha_n=0.1$. In the simulation we 
remove the heat generated by dissipative effects.}   
\end{figure}

 We first consider the evolution of a Gaussian density profile in ideal 
hydrodynamics. The initial condition is given by equ.~(\ref{n_0_barx})
where we have chosen $E_0/E_F=1$ and $\bar{n}_0=1$ (the ideal evolution
is independent of the number of particles). The aspect ratio of the 
cloud is $\lambda=0.045$. In Fig.~\ref{fig_prof_tr} we show the evolution 
of the density and the velocity. The points are the results of the numerical 
solution using VH1 and the lines are semi-analytic results based on solving 
the coupled set of ordinary differential equations~(\ref{b_i_ode}). The 
numerical calculation was performed on a fairly coarse grid with a grid
spacing $\Delta \bar{x}=0.2$. We observe that the numerical calculation 
is nevertheless very accurate. 

 The lower panel of Fig.~\ref{fig_prof_tr} shows that the velocity 
field tracks the linear behavior of the analytic solution only up to 
some maximum distance which slowly grows with time. The turnover of 
the velocity field is related to the fact that we impose a minimum 
density and pressure ($10^{-15}$ of the initial central density and 
pressure). Beyond the point at which the minimum pressure is reached 
there are no pressure gradients and therefore no acceleration. The 
sharp discontinuity in the velocity does not lead to numerical problems. 
A smooth turnover of the velocity field can be achieved by considering 
slightly modified initial conditions. If we solve the equation of
hydrostatic equilibrium in a potential which is harmonic at short 
distances, but grows as $V\sim |\vec{x}|^\alpha$ with $\alpha<1$ 
at large distances, then the velocity field of the expanding cloud
will go to zero smoothly as $|\vec{x}|\to\infty$. 

 We next consider the dissipative evolution of a Gaussian density 
profile. We take the shear viscosity to be of the form $\eta=\alpha_n
n$ with a constant $\alpha_n$. In order to compare with the solution 
discussed in Sect.~\ref{sec_exact} we include a sink in the equation
of energy conservation as defined in equ.~(\ref{heat_sink}). Note 
that we can write the divergence of the dissipative energy current as
\be 
\nabla_i (\delta\jmath^{\epsilon}_i) = 
\nabla_i \left( v_j\delta\Pi_{ij} \right) = 
-\frac{\eta}{2}\left(\sigma_{ij}\right)^2
 + v_i\nabla_j \delta\Pi_{ij}\, .
\ee
The first terms corresponds to viscous heating and the second term 
describes the work done by viscous forces. Adding a heat sink implies
that we only keep the effect of the work term. 

 The evolution of the kinetic, potential, and total energy for 
$\bar{\alpha}_n=0.1$ and $E_0/E_F=1$ are shown in Fig.~\ref{fig_e_t_work}. 
The solid lines show the ideal evolution determined by equ.~(\ref{b_i_ode})
and the dashed lines show the dissipative result given by the solution of 
equ.~(\ref{ns_mom_1},\ref{ns_mom_2}) for $\beta=\frac{2}{3}\bar{\alpha}_n
(E_F/E_0)=0.066$. The data points 
come from a numerical calculation based on the dissipative version of 
VH1. We observe that the data agree very well with the analytical 
solution. The main difference between the dissipative and the 
ideal solution is that a fraction of the kinetic energy is converted
to heat. The heat is absorbed by the sink and lost to the system. 
The evolution of the internal energy is only affected indirectly,
via the effect of dissipation on the evolution of the radius 
of the system.

\begin{figure}[t]
\bc\includegraphics[width=10cm]{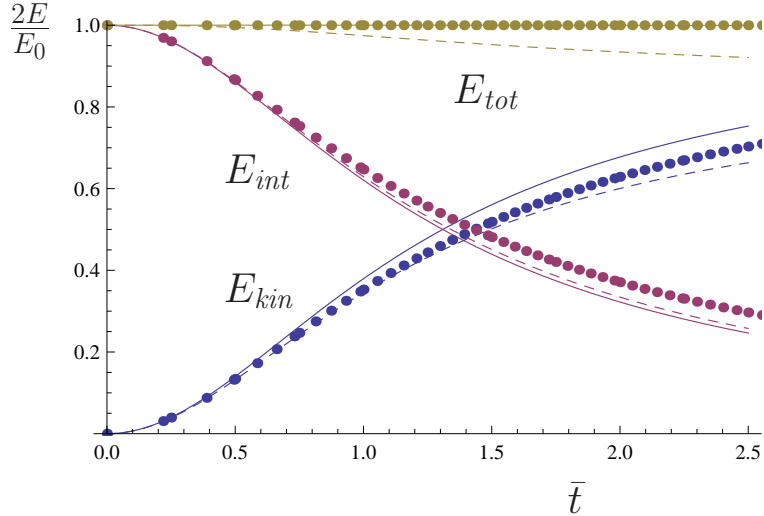}\ec
\caption{\label{fig_e_t_full}
Energy of the expanding gas cloud as a function of time. We show the
kinetic, internal, and total energy. The solid lines show the
analytical result for the ideal evolution, and the dashed lines 
show the analytical result without reheating for $\beta=0.066$. 
The data points show the result of a numerical calculation with 
$E_0/E_F=1$ and $\bar\eta=\bar\alpha_n \bar{n}$ with $\bar\alpha_n=0.1$. 
The simulation includes all dissipative terms in the energy current.}   
\end{figure}

\subsection{Effects of reheating}
\label{sec_reh}
 
 The evolution of the energy in a complete simulation, including the 
effects of reheating, is shown in Fig.~\ref{fig_e_t_full}. The system
parameters are the same as in the previous section. We observe that 
the total energy is conserved to a very good accuracy. Reheating 
increases the internal energy as compared to the result in ideal 
hydrodynamics. Hydrodynamic evolution converts the added internal 
energy into kinetic energy. This implies that reheating leads to 
reacceleration. 

\begin{figure}[t]
\bc\includegraphics[width=10.0cm]{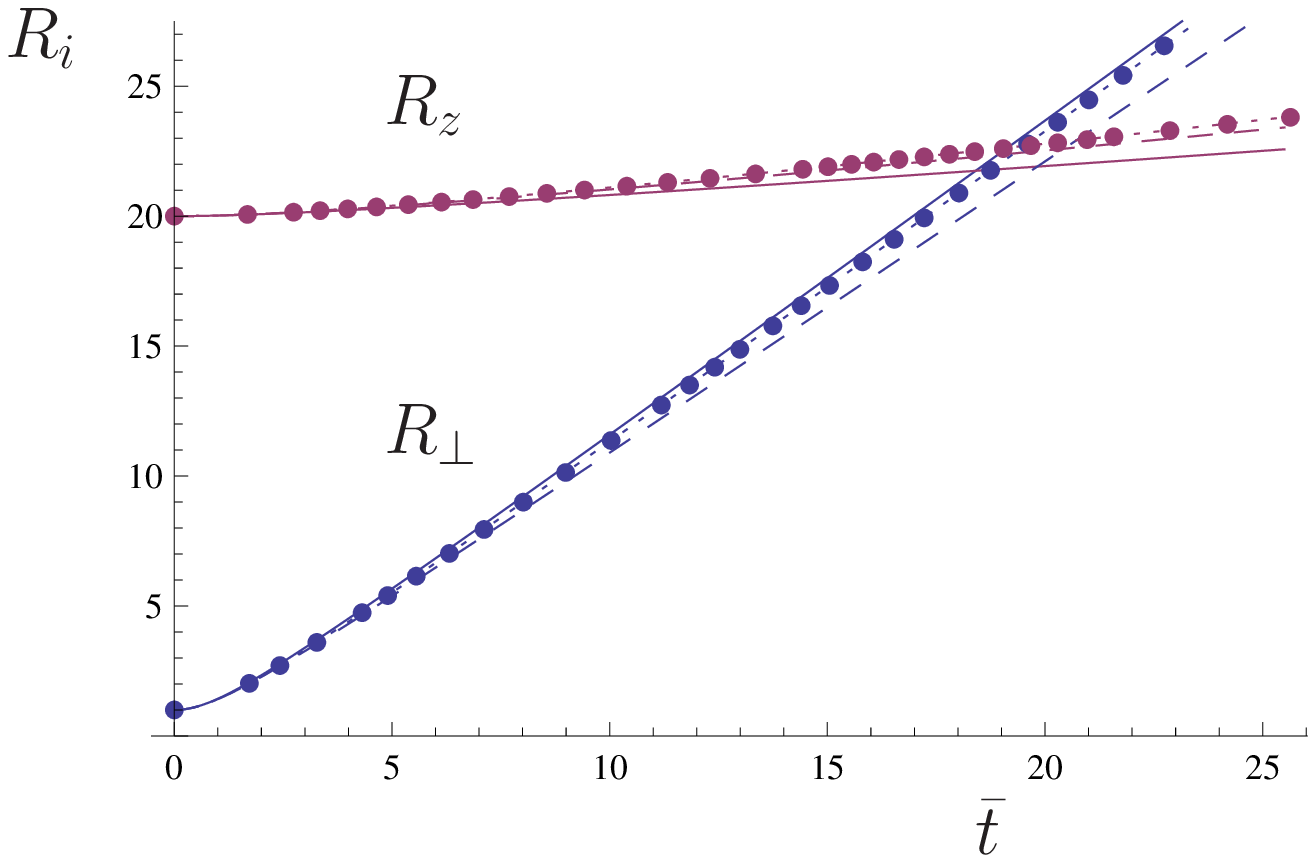}\ec
\bc\includegraphics[width=10.0cm]{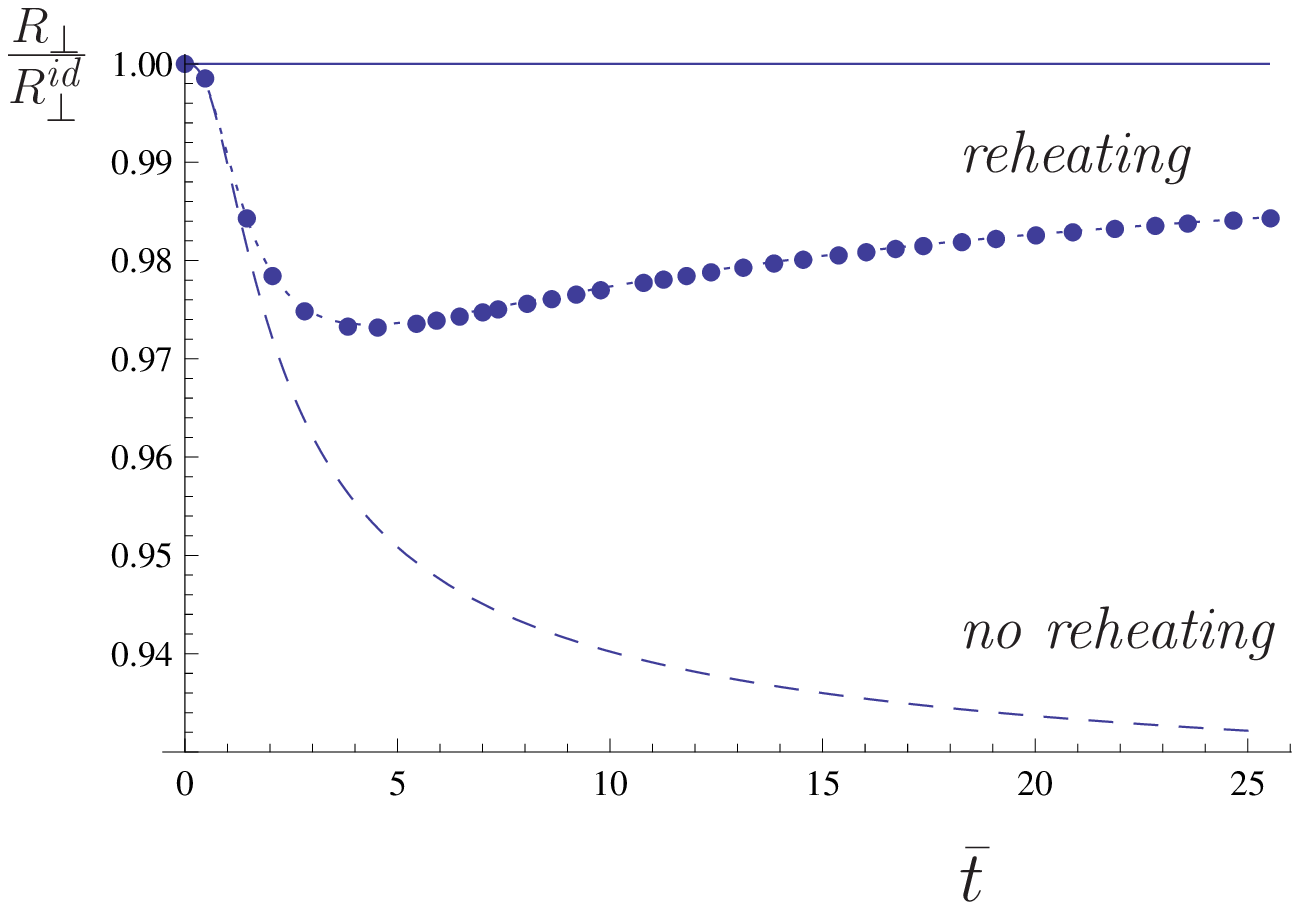}\ec
\caption{\label{fig_r_t_full}
The top panel shows the transverse and longitudinal scale factors 
of the expanding gas cloud as a function of time. The initial density
has a Gaussian profile with $\lambda=0.05$ and $E_0/E_F=1$. The viscosity 
is of the form $\bar\eta=\bar\alpha_n \bar{n}$ with $\bar\alpha_n=0.1$. 
The solid lines show the analytical result for the ideal evolution, the 
dashed lines correspond to the dissipative solution without reheating, 
and the dotted line is the approximate solution described in 
Sect.~\ref{sec_app}. The points are from a numerical calculation.
The bottom panel shows the ratio of the transverse scale factor 
over the result in ideal hydrodynamics for the calculations shown 
in the top panel.  }   
\end{figure}

 This is shown in more detail in Fig.~\ref{fig_r_t_full}. The 
upper panel shows the time evolution of the Gaussian radii $R_\perp(t)$ 
and $R_z(t)$. We have normalized $R_\perp(0)=1$ and $R_z(0)=1/\lambda$
so that $R_\perp(T)=R_z(t)$ corresponds to an aspect ratio of one.
The solid lines show the result in ideal hydrodynamics, the dashed 
lines show the result without reheating, the dotted line shows 
the approximate solution including reheating discussed in 
Sect.~\ref{sec_app}, and the data points are obtained from a 
numerical calculation. The lower panel shows the ratio $R_\perp(t)
/R^{\it id}_\perp(t)$ where $R^{\it id}_\perp(t)$ is the transverse
size in ideal hydrodynamics. We observe that the two radii initially
track the prediction of the calculation without reheating: Viscosity 
slows down the expansion in the short direction, and accelerates the 
system in the longitudinal direction. At later times reheating leads 
to an acceleration in both directions. The lower panel of 
Fig.~\ref{fig_r_t_full} shows that the transverse size almost 
goes back to the prediction of ideal hydrodynamics. For a shear 
viscosity which is linear in the density this behavior is very well 
described by the linear force model discussed in Sect.~\ref{sec_app}.

\begin{figure}[t]
\bc\includegraphics[width=10.0cm]{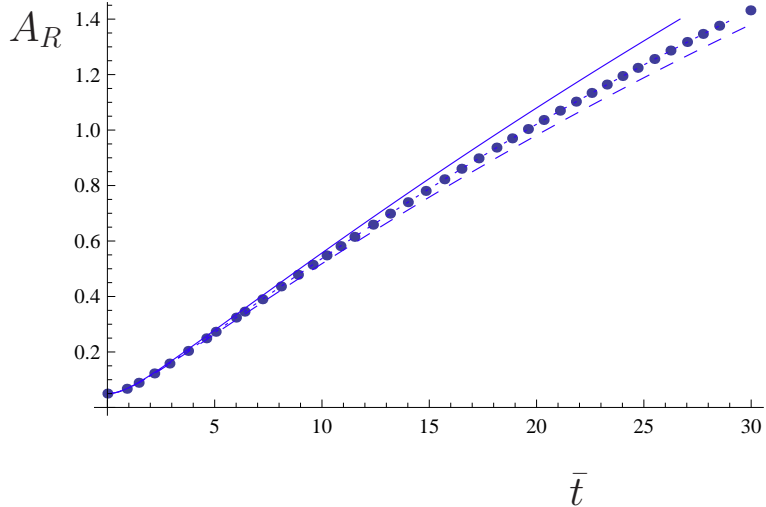}\ec
\caption{\label{fig_aspect}
This figure shows the time evolution of the aspect ratio $A_R=
\lambda R_\perp(t)/R_z(t)$. The solid lines show the analytical 
result for the ideal evolution, the dashed lines correspond to 
the dissipative solution without reheating, and the dotted line 
is the approximate solution described in Sect.~\ref{sec_app}. 
The points are from a numerical calculation. All curves correspond
to a Gaussian initial condition, and the viscosity is of the form 
$\bar\eta=\bar\alpha_n\bar{n}$ with $\bar\alpha_n=0.1$.}   
\end{figure}

 Fig.~\ref{fig_aspect} shows that even if reheating is taken into
account shear viscosity leads to significant effects in the evolution 
of the aspect ratio $A_R=\lambda R_\perp/R_z$. The shape of $A_R(t)$ 
is similar in the model without reheating and the numerical simulation, 
but the magnitude of the dissipative effect is about a factor 2 smaller 
if reheating is taken into account. We observe that shear viscosity leads 
to a characteristic bending of $A_R(t)$ in the regime $A_R\sim 1$. In 
ideal hydrodynamics acceleration takes place at early times $\bar{t}
\lsim 3$. Dissipative forces and reheating lead to longitudinal 
acceleration which occurs on a much longer time scale. Observing 
this behavior not only constrains the value of the shear viscosity, 
it also demonstrates that the systems continues to behave hydrodynamically 
even at very late times. 

\begin{figure}[t]
\bc\includegraphics[width=10.0cm]{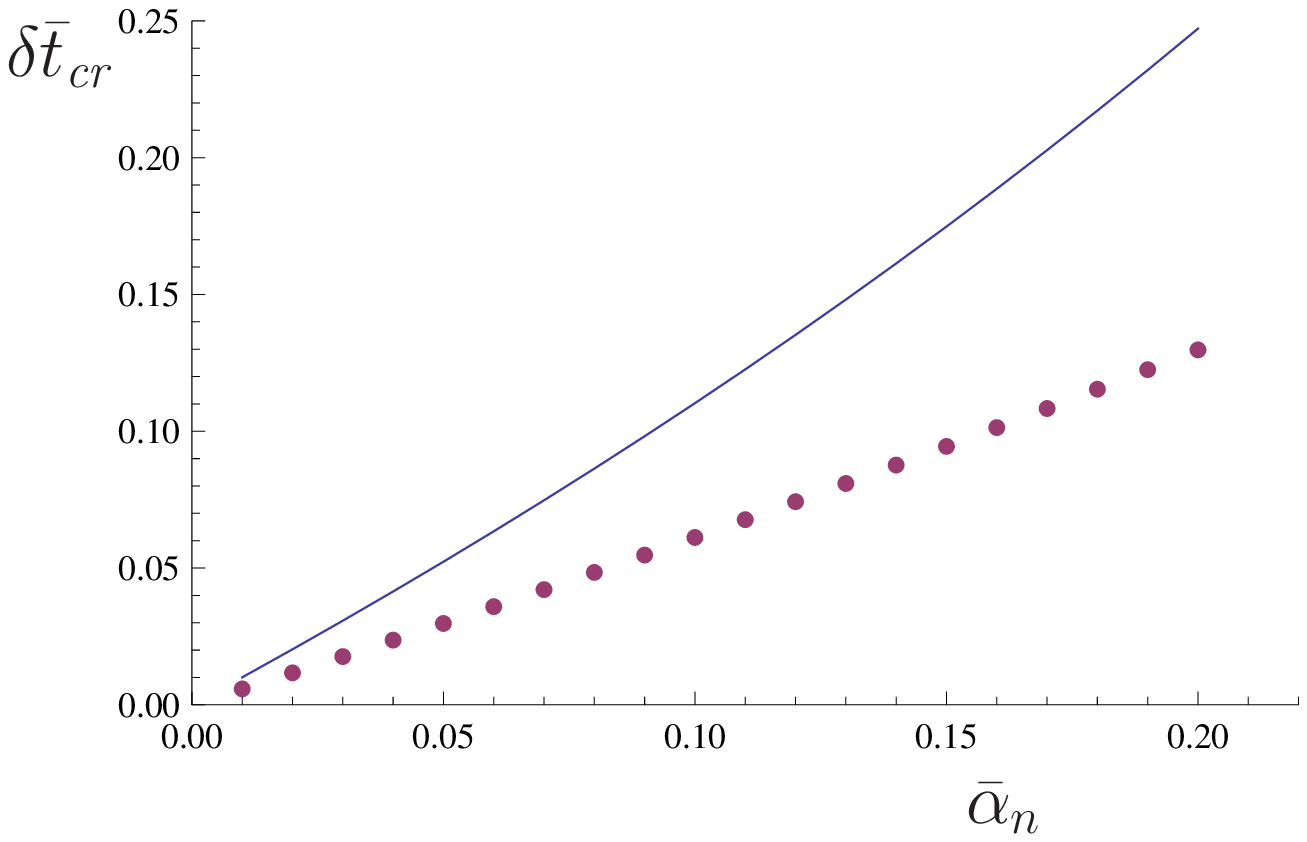}\ec
\bc\includegraphics[width=10.0cm]{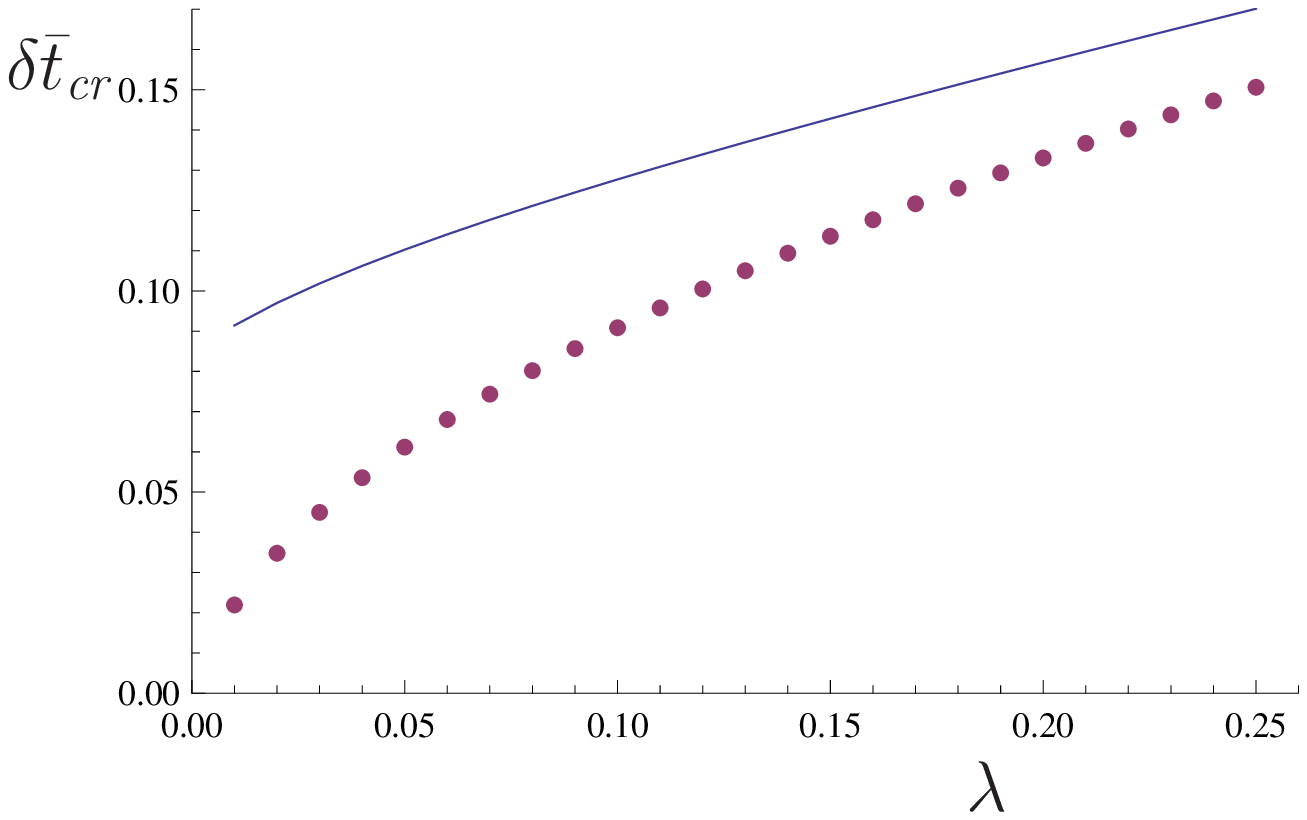}\ec
\caption{\label{fig_del_t}
This figure shows the viscous correction to the crossing time $t_{\it cr}$
defined by $A_R(t_{\it cr})\equiv 1$. The top panel shows $\delta 
\bar{t}_{\it cr}$ as a function of the viscosity $\bar\alpha_n$, 
where $\bar\eta=\bar{\alpha}_n\bar{n}$, for $E_0/E_F=1$ and a fixed 
initial aspect ratio $\lambda=0.05$. The solid line is the result 
of the dissipative calculation without reheating and the points 
are from a numerical calculation. The bottom panel shows $\delta 
\bar{t}_{\it cr}$ as a function of $\lambda$ for a fixed viscosity 
$\bar\alpha_n=0.1$ }   
\end{figure}
 
 The effect of shear viscosity on the evolution of $A_R(t)$ can 
be quantified in terms of the ``crossing time'' $t_{\it cr}$ defined
by $A_R(t_{\it cr})\equiv 1$. Viscosity leads to a shift $\delta t_{\it cr}$
in the crossing time as compared to ideal hydrodynamics. In 
Fig.~\ref{fig_del_t} we show $\delta \bar{t}_{\it cr}$ as a function of 
$\bar\alpha_n$ and $\lambda$. Ideal hydrodynamics predicts that 
$t_{\it cr}=\sqrt{\gamma}/(\lambda\omega_\perp)$ with $\gamma=
2/3$. This result is correct in the limit $\lambda\ll 1$ up to 
higher order corrections in $\lambda$. Neglecting the effects of 
reheating the correction to the crossing time is \cite{Schaefer:2009px}
\be
 \left( \frac{\delta t}{t}\right)_{\it cr} = 
 1.16\, \frac{\langle \alpha_n\rangle}{(3N\lambda)^{1/3}}
      \frac{1}{(E_0/E_F)}\, , 
\ee
where $\langle\alpha_n\rangle$ is the average of $\alpha_n$ over the 
initial density of the trap. The change in $t_{\it cr}$ if reheating 
is included is shown in Fig.~\ref{fig_del_t}. In the upper panel 
we show the dependence of $\delta \bar{t}_{\it cr}$ on $\bar{\alpha}_n$. 
We observe that the effect remains linear if reheating is included, 
but that the sensitivity of $\delta \bar{t}_{\it cr}$ to $\bar{\alpha}_n$ 
is reduced by about a factor of 2. The lower panel shows that the
correction factor depends on the geometry. If reheating is included
then the sensitivity of $\delta \bar{t}_{\it cr}$ to the shear viscosity
becomes very small in the limit of strongly deformed traps ($\lambda\to
0$). This is related to the fact that in this limit all internal
energy is converted to transverse motion, irrespective of whether 
or not there is dissipation. 

\begin{figure}[t]
\bc\includegraphics[width=10.0cm]{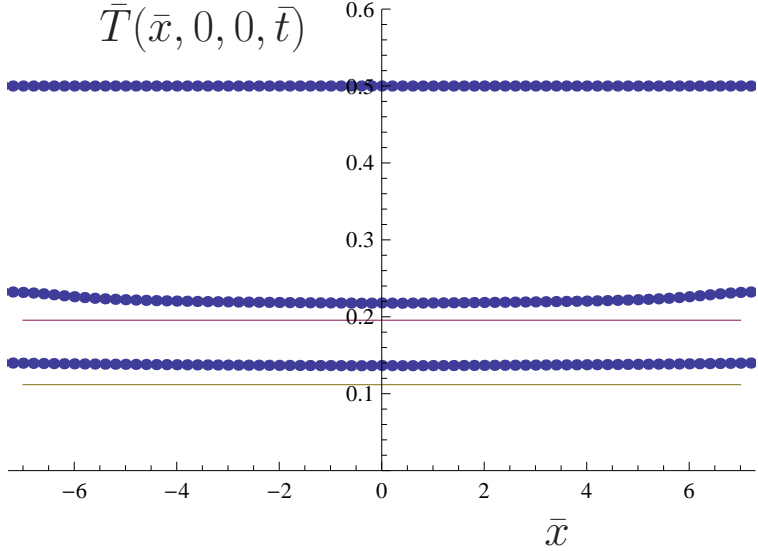}\ec
\caption{\label{fig_temp}
This figure shows the evolution of the temperature profile in viscous 
hydrodynamics. The data points show the temperature $\bar{T}(\bar{x},
0,0,\bar{t})$ determined in a numerical simulation with $\bar{\alpha}
=0.1$ at several different times $\bar{t}=0,1.68,2.70$. The lines are 
the result in ideal hydrodynamics. }   
\end{figure}

\begin{figure}[t]
\bc\includegraphics[width=10.0cm]{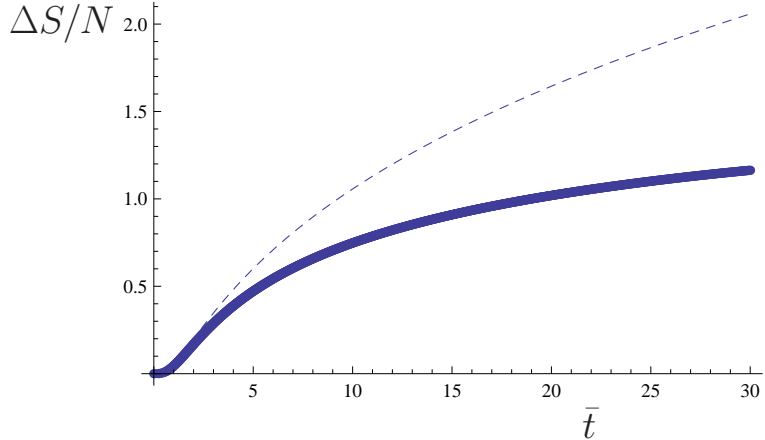}\ec
\caption{\label{fig_entr}
This figure shows the evolution of the produced entropy per particle 
$\Delta S/N$ as a function of time. The thick line shows the result 
in a numerical simulation with $E_0/E_F=1$ and $\bar{\alpha}_n=0.1$ 
and the thine line is the entropy removed by the heat sink in the 
model discussed in Sect.~\ref{sec_exact}.}   
\end{figure}

 Fig.~\ref{fig_temp} shows the effect of reheating on the temperature 
profile of the cloud. The solid line shows the temperature profile at 
different times during the ideal evolution, and the data points come 
from a numerical simulation with $\bar\alpha_n=0.1$. The increase in 
the average temperature relative to the result in ideal hydrodynamics 
is due to reheating combined with a decrease in the expansion 
rate. In our simulation we have used a spatially constant $\alpha_n$
and the high temperature equation of state $P=nT$. In this case
both the dissipated heat and the specific heat are proportional 
to the density. As a consequence the cloud remains isothermal 
to a fairly good accuracy. 

 Fig.~\ref{fig_temp} shows the change in the entropy per particle 
during the evolution of the system.  The data points show the result 
of a simulation using an ideal gas equation of state with 
$E_0/E_F=1$ and a shear viscosity $\bar\eta=\bar\alpha_n \bar{n}$
with $\bar\alpha_n=0.1$. For the ideal gas equation of state 
we can compute the entropy using the Sackur-Tetrode formula.
The dashed line shows the entropy absorbed by the the heat sink 
for a calculations with no reheating of the gas. In this case
the produced entropy scales asymptotically as \cite{Schaefer:2009px}
\be 
\label{del_S}
\frac{\Delta S}{N} \simeq 4\left(\frac{2}{3}\right)^{1/3}  
 \frac{\langle \bar\alpha_n\rangle}{(T_0/T_F)} 
  \left(\omega_\perp t\right)^{1/3}\, .
\ee 
The full simulation tracks the results without reheating
very well for $(\omega_\perp t)\lsim 3$. At later times the 
full simulation produces less entropy then the model without
reheating. However, we still find that the total entropy 
continues to grow as $t\to\infty$. Numerically, we find
that the asymptotic behavior is well described by 
$(\Delta S)/N\sim (\omega_\perp t)^{1/6}$. 

\begin{figure}[t]
\bc\includegraphics[width=10.0cm]{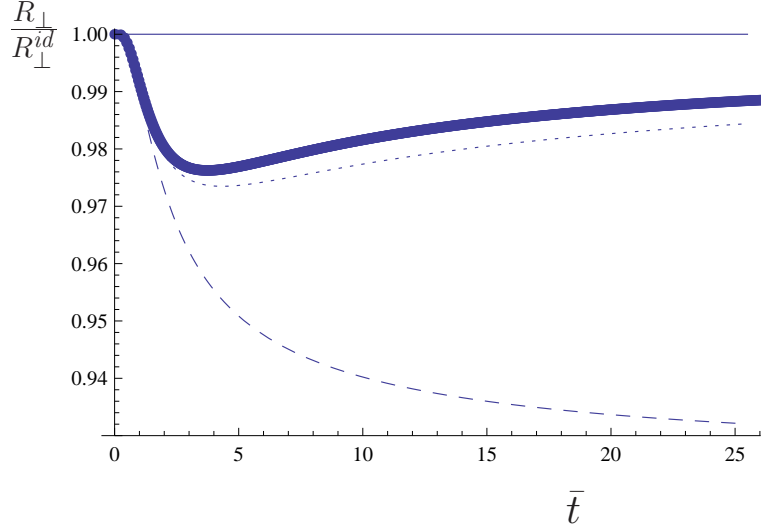}\ec
\caption{\label{fig_rperp_qu}
This figure shows the ratio of the transverse scale factor in 
dissipative hydrodynamics over the result in ideal hydrodynamics.
The data points are from a simulation with $\eta\sim n^2$ and 
$\langle\bar{\alpha}_n\rangle =0.1$. The lines show the same 
calculations as in Fig.~\ref{fig_r_t_full}. }   
\end{figure}

\subsection{Dependence on the equation of state and the functional
form of the shear viscosity}
\label{sec_eos_eta}

 There are several arguments that indicate that the evolution of 
the system is not very sensitive to the equation of state $P(n,T)$ 
as long as the universal relation $P=\frac{2}{3}{\cal E}$ is satisfied. 
In Sect.~\ref{sec_exact} we showed that the exact solution of Euler's 
equation is independent of the equation of state. We also showed that 
the equations of dissipative hydrodynamics given in 
equ.~(\ref{ns_force},\ref{q_force}) are independent of the equation of 
state as long as the velocity field remains exactly linear. In this 
section we will study the dependence on the equation of state using 
numerical simulations of the complete hydrodynamic equations. We will 
compare the results obtained using the ideal gas equation of state $P=nT$ 
and the equation of state described in Appendix \ref{sec_eos}. We consider 
a temperature $T=0.25T_F$, close to the superfluid phase transition, 
where the deviation of the experimental equation of state from the ideal 
gas equation is largest. For the ideal gas equation of state we have 
$(E_0/E_F)=3(T/T_F)=0.6$. We choose $\bar\alpha_n= 0.06$ so that $\beta=
0.066$. For the experimental equation of state we find $E_0/E_F=0.785$ 
and we set $\bar\alpha_n = 0.0785$ to keep $\beta$ fixed. We find that 
the effect of the equation of state on the change in $t_{\it cr}$ is 
smaller than the accuracy of our calculation, $(\delta t_{\it cr}
(P^{\it id})-\delta t_{\it cr}(P^{\it ex}))/t_{\it cr}<10^{-3}$.

\begin{figure}[t]
\bc\includegraphics[width=10.0cm]{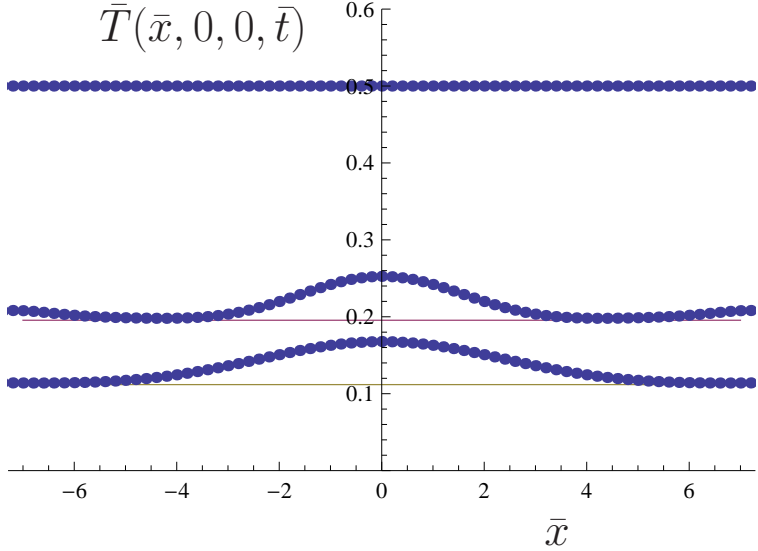}\ec
\caption{\label{fig_temp_qu}
This figure shows the evolution of the temperature profile in viscous 
hydrodynamics. The data points show the temperature $\bar{T}(\bar{x},
0,0,\bar{t})$ determined in a numerical simulation with $\eta\sim n^2$
and $\langle\bar{\alpha}\rangle =0.1$ at several different  times 
$\bar{t}=0,1.65,2.66$. The lines are the result in ideal hydrodynamics.}   
\end{figure}

 We have also studied the dependence of dissipative effects on the 
functional form of the shear viscosity. The approximate solutions
discussed in Sections \ref{sec_exact} and \ref{sec_app} suggest 
that dissipative effects depend only on the trap average $\langle 
\alpha_n\rangle$, see equ.~(\ref{beta_def}). In the following we 
will test this idea by comparing calculations with $\alpha_n\sim
{\it const}$, corresponding to $\eta\sim n$, and $\alpha_n\sim 
n/(mT)^{3/2}$, which implies $\eta\sim n^2/(mT)^{3/2}$. We write 
$\eta=\eta_2 n^2/(mT)^{3/2}$ and fix $\eta_2$ from $\langle
\alpha_n\rangle$. For a Gaussian profile 
\be 
\eta_2 = 24\pi^{3/2} \langle \alpha_n \rangle 
 \left( \frac{T}{T_F}\right)^3\, . 
\ee
Fig.~\ref{fig_rperp_qu} shows the evolution of the transverse 
radius for $\langle\bar\alpha_n\rangle=0.1$. The lines are the same 
as in Fig.~\ref{fig_r_t_full}. We observe that the calculations with 
$\eta\sim n$ and $\eta\sim n^2$ are very similar for $(\omega_\perp
t)\lsim 3$. At later times the non-linear dependence of $\eta$
on $n$ leads to some extra acceleration. Fig.~\ref{fig_temp_qu}
shows the corresponding temperature profiles. We observe that 
for $\eta\sim n^2$ reheating takes place predominantly near the
center of the cloud. This leads to an outward temperature 
gradient which is the source of the additional acceleration. 
Quantitatively, the difference between the $\eta\sim n$ and 
$\eta\sim n^2$ is about 25\%,  
\be
\frac{\delta t_{\it cr}(\eta\sim n)
          -\delta t_{\it cr}(\eta\sim n^2)}
     {\delta t_{\it cr}(\eta\sim n)} = 0.264\,. 
\ee

\subsection{Rotating solutions}
\label{sec_rot}

 In this section we study the time evolution of a rotating cloud. 
We take the initial velocity profile to be of the form $\vec{v} = 
\alpha\vec{\nabla}(xz)$ with $\alpha(0)=\omega_{\it rot}=0.4\omega_z$ 
\cite{Clancy:2007}. We have checked that the results are unaffected 
by taking the initial profile to be of the form $\vec{v} = \Omega\hat{y}
\times\vec{x}$. The reason is that for a strongly deformed cloud the 
initial momentum density $\rho\vec{v}$ is essentially the same for 
irrotational or rigid initial conditions. 

\begin{figure}[t]
\bc\includegraphics[width=10.0cm]{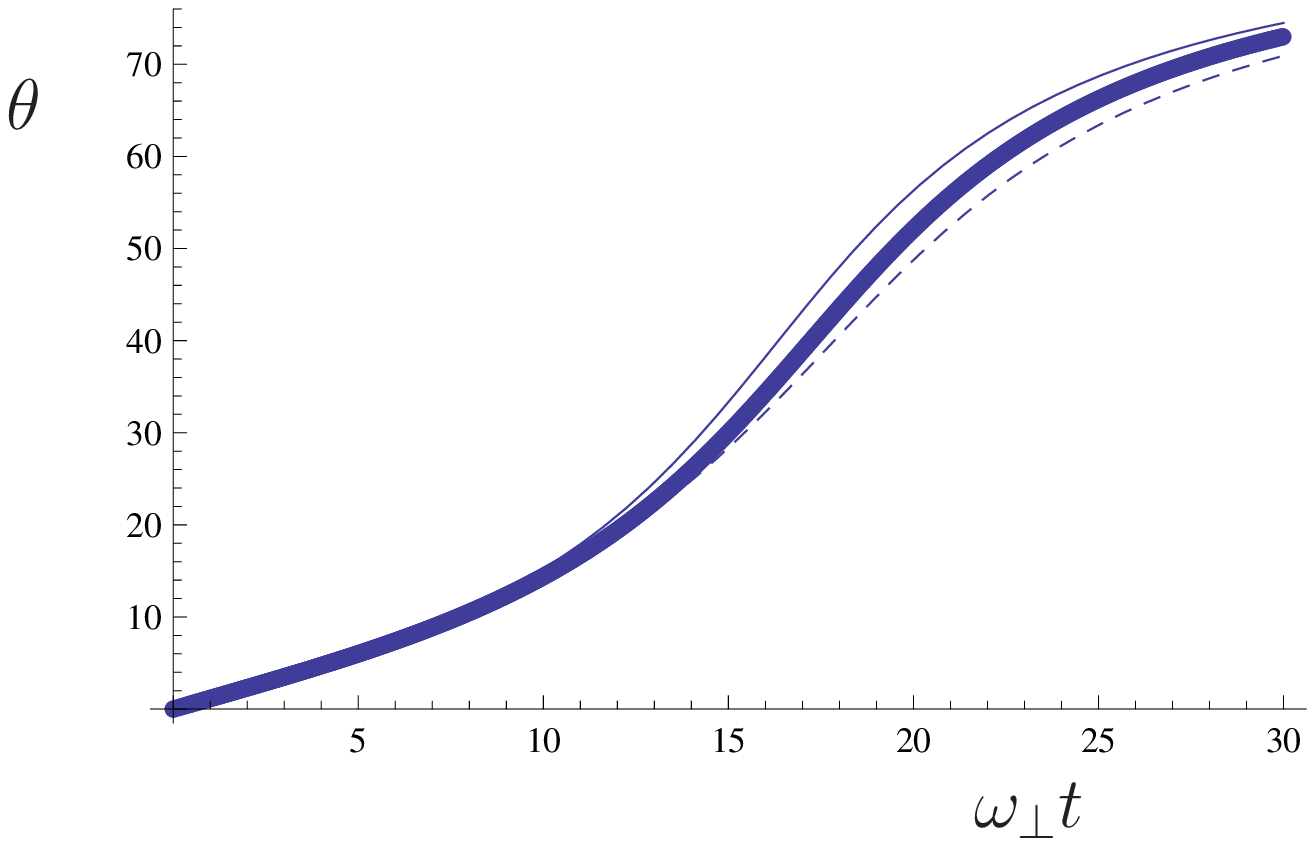}\ec
\bc\includegraphics[width=10.0cm]{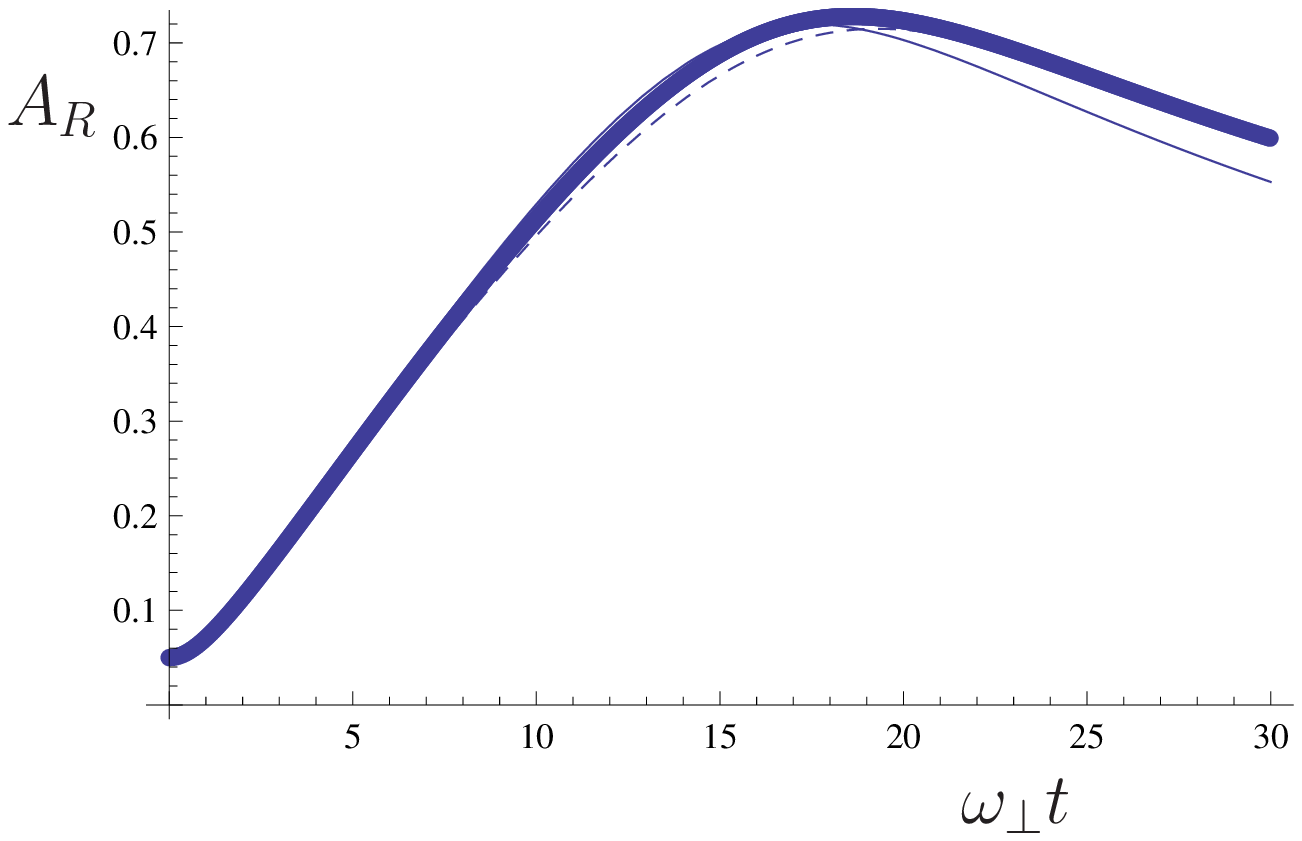}\ec
\caption{\label{fig_rot}
This figure shows the time evolution of the angle (top panel) and the 
aspect ratio (bottom panel) of a rotating cloud as a function of time. 
The initial density profile is Gaussian, and the initial flow profile
is an irrotational flow with $\omega_{\it rot}=0.4\omega_z$. The thin 
line is the result in ideal hydrodynamics, the dashed line is the result in 
dissipative hydrodynamics with a heat sink, and the thick line is 
result of a numerical calculation with $\bar\alpha=0.1$. }   
\end{figure}

 We determine the angle of the major axis and the aspect ratio of 
the cloud. The angle is related to the Gaussian radii by 
\be 
\tan(2\theta) = 
 \frac{2\langle xz \rangle}{\langle z^2\rangle -\langle x^2\rangle}\, , 
\ee
and the aspect ratio is given by
\be 
A_R = \left\{ 
 \frac{\langle x^2\rangle +\langle z^2\rangle
    +\left[\left(\langle z^2\rangle -\langle x^2\rangle\right)^2 
              +4\langle xz\rangle^2 \right]^{1/2}}
      {\langle x^2\rangle +\langle z^2\rangle
    -\left[\left(\langle z^2\rangle -\langle x^2\rangle\right)^2 
              +4\langle xz\rangle^2 \right]^{1/2}}
\right\}^{1/2}\, . 
\ee
Our results are shown in Fig.~\ref{fig_rot}. The solid line is 
the result in ideal hydrodynamics. A good approximation to the 
evolution of the angle in ideal hydrodynamics is 
\be
\tan(2\theta) = - \frac{a\lambda^2b_\perp^2b_z^2}
  {b_\perp^2-\lambda^2 b_z^2}\, , 
\ee
where $b_\perp,b_z$ are the scale parameters for the pure 
expansion (without rotation) and 
\be 
\label{a_as}
a(t)  \simeq \left\{ \begin{array}{cl}
  -\frac{2\omega_{\it rot}t}{\lambda^2} \;\; 
                    & \omega_\perp t\ll 1 \, , \\
  - \frac{\gamma\omega_{\it rot}}{\lambda^2\omega_\perp^2 t}
                     & \omega_\perp t\gg 1 \,  ,
\end{array}\right.
\ee
with $\gamma=2/3$. This result shows that the angle goes through 45 
degrees at the same time at which the expanding system reaches an 
aspect ratio of 1. 

 Fig.~\ref{fig_rot} shows the time evolution of the angle and the 
aspect ratio. The solid line shows the result in ideal hydrodynamics, 
the dashed line shows the result in dissipative hydrodynamics neglecting 
reheating, and the data points are from a numerical simulation with 
$\bar{\eta}= \bar{\alpha}_n\bar{n}$ and $\bar{\alpha}_n=0.1$. We 
observe that the effect of reheating in rotating clouds is similar
to the effect in non-rotating systems. Reheating accelerates the 
system and reduces dissipative corrections. This effect can be 
quantified in terms of the time $t_{45^\circ}$ at which the angle 
of the major axis passes through $45^\circ$ (angular momentum 
conservation combined with the approximately irrotational nature 
of the flow implies that the aspect ratio never reaches the value 
1). We find that, within the accuracy of our calculation, the 
dissipative correction to $t_{45^\circ}$ is equal to the dissipative 
correction to the crossing time (see Sect.~\ref{sec_reh}), $\delta 
t_{45^\circ}=\delta t_{\it cr}$. This implies, in particular, that 
earlier estimates of the shear viscosity based on calculations that 
do not take into account reheating have to be a corrected by a 
factor $\sim 2$ \cite{Thomas:2009zz,Schaefer:2009px}.

\section{Conclusions and outlook}
\label{sec_sum}

 In this work we studied the expansion dynamics of a dilute 
Fermi gas at unitarity in the framework of dissipative hydrodynamics. 
Our main goal was to study whether one can extract the shear 
viscosity from expanding systems. This is not immediately obvious, 
because in an expanding system all internal energy is eventually
converted into kinetic energy, irrespective of whether there is 
dissipation or not. 

 We find that shear viscosity does lead to characteristic effects 
in the expansion dynamics. Shear viscosity causes a characteristic 
curvature in the time evolution of the aspect ratio $A_R(t)$ of 
the cloud. In ideal hydrodynamics internal energy is converted to
kinetic energy very quickly, over a time period $(\omega_\perp t)
\lsim 3$. After this time $A_R(t)$ is essentially linear. In 
dissipative hydrodynamics energy stored in the transverse motion 
is converted into longitudinal kinetic energy, and the longitudinal 
expansion takes place on a much longer time scale. As a result
$A_R(t)$ exhibits a characteristic curvature at times as large
as $(\omega_\perp t)\simeq \lambda^{-1}\simeq 25$. This effect was
recently observed by Cao et al.~\cite{Cao:2010}, which shows that 
dissipative hydrodynamics is indeed valid at $(\omega_\perp t)\simeq 
25$. This is a remarkable discovery, because during the evolution
the density drops by a factor $\lambda^{-2}\sim 10^3$.

 We also find that a quantitative description of the dependence 
of $A_R(t)$ and other observables on the shear viscosity has to 
include reheating. For a cloud with an aspect ratio of $25$ the 
extracted shear viscosity is about a factor 2 too small if 
reheating is neglected. This affects the estimates presented 
in \cite{Clancy:2008,Thomas:2009zz,Schaefer:2009px} but not 
the recent work of Cao et al.~\cite{Cao:2010}. Reheating also 
does not affect estimates of the shear viscosity based on the 
damping of collective modes \cite{Schafer:2007pr,Turlapov:2007}.

 We showed that a determination of the shear viscosity does not 
require an accurate knowledge of the equation of state $P(n,T)$. 
The only important aspect of the equation of state is the universal 
relation $P=\frac{2}{3}{\cal E}$. We also studied the dependence of 
viscous effects on the functional form of $\eta(n,T)$. We find that 
to first approximation the expansion dynamics constrains the cloud 
average of the shear viscosity. In this approximation the universal 
function $\alpha_n(mT/n^{2/3})$ can be determined by extracting 
$\langle\alpha_n\rangle$ as a function of $T/T_F$ from data, and
then inverting equ.~(\ref{alpha_av}). This only requires knowledge 
of the initial density profile. The result can be used as input for a 
more accurate determination based on full hydrodynamics.

 There are several issues that remain to be studied. The most 
important problem has to do with the breakdown of hydrodynamics 
in the dilute corona of the cloud. In the low density, high 
temperature limit the shear viscosity can be reliably computed. 
The result shows that the shear viscosity is independent of 
the density, $\eta\sim (mT)^{3/2}$. This implies that the 
total amount of heat dissipated by the dilute tail of the 
density distribution is infinite. We have previously argued that 
this problem can be resolved by taking into account the fact that
the dissipative contribution to the stress tensor relaxes
to the Navier-Stokes form on a time scale which is proportional
to the density of the system \cite{Bruun:2007,Schaefer:2009px}.
In kinetic theory we expect that $\tau_R\partial_t(\delta\Pi_{ij})
=(\eta\sigma_{ij}-\delta\Pi_{ij})$ where the relaxation time 
is given by $\tau_R=\eta/(nT)$. This implies that in the dense
regime the shear viscosity relaxes to its equilibrium value 
on a time scale that is fast compared to the time scale of 
the hydrodynamic expansion, but in the dilute regime dissipation 
is governed by an effective viscosity which is proportional to 
the density. 

 This idea can be implemented by using an effective $\langle
\alpha_n\rangle$ in solving the equations of dissipative 
hydrodynamics \cite{Schaefer:2009px,Cao:2010}. It is 
clearly preferable, however, to include the effects of 
finite relaxation time by including higher derivative 
terms in the equations of dissipative fluid dynamics
(this is known as 2nd order, or Burnett, hydrodynamics), 
or by coupling the hydrodynamic description to kinetic
theory.

 Acknowledgments: This work was supported in parts by the US 
Department of Energy grant DE-FG02-03ER41260. We would like to 
thank Clifford Chafin and John Thomas for useful discussions 
and John Blondin for help with VH1.

\appendix
\section{Equation of state}
\label{sec_eos}

\begin{figure}[t]
\bc\includegraphics[width=10cm]{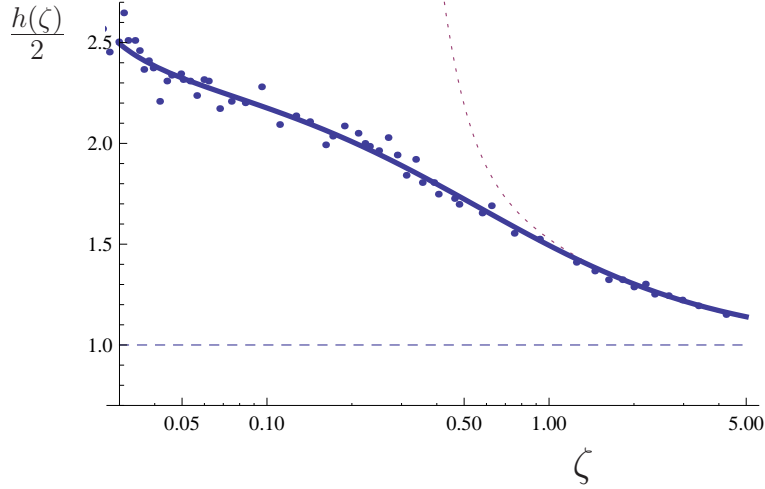}\ec
\caption{\label{fig_eos}
Equation of state of the unitary Fermi gas in the normal phase. 
In this figure we show the function $h(\zeta)$, where $\zeta$ is
the fugacity. The data points are from Nascimbene et al., the thick
solid line shows the parameterization discussed in the text, the dashed
line is the non-interacting gas result $h=1$, and the dotted line shows
the second order Virial expansion.}   
\end{figure}
\begin{figure}[t]
\bc\includegraphics[width=10cm]{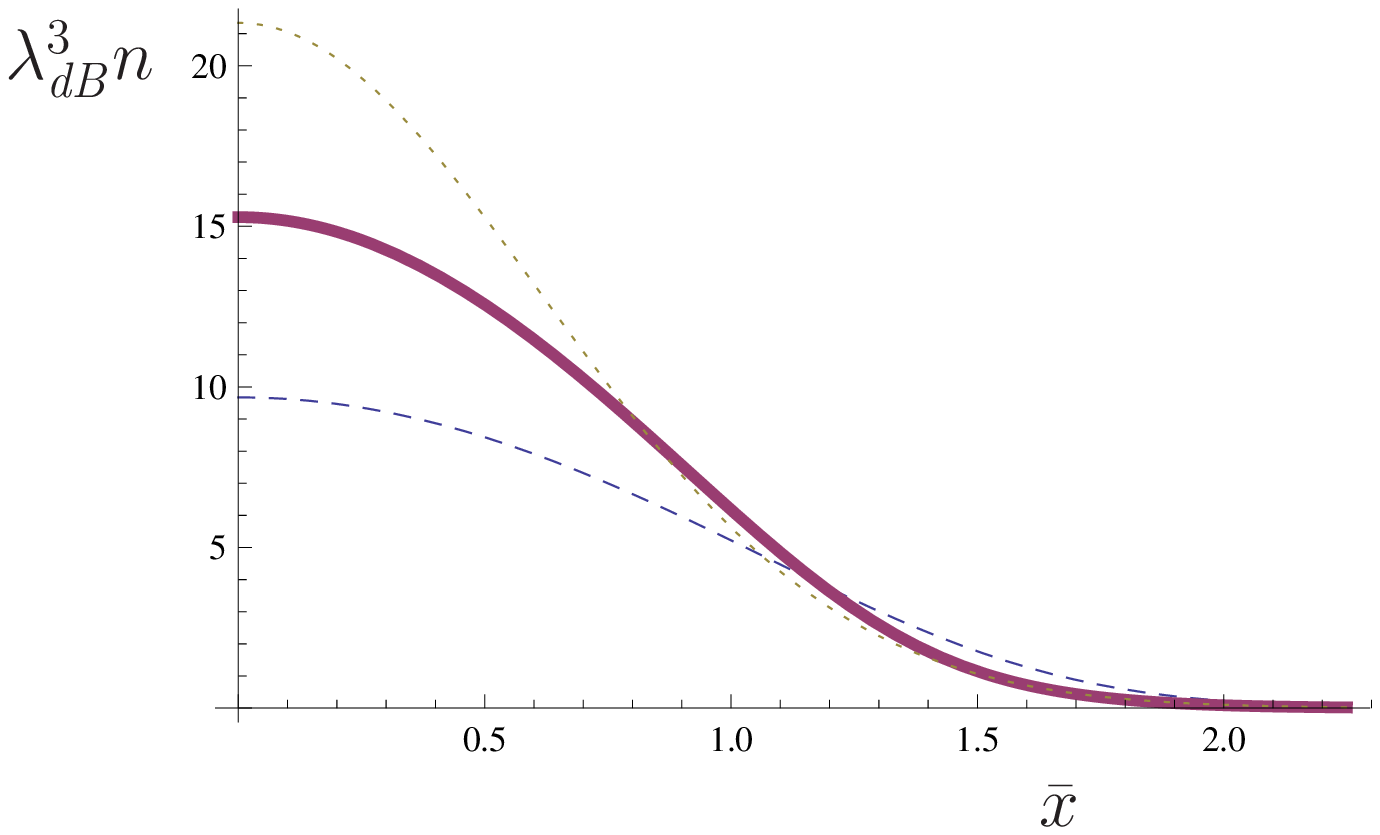}\ec
\caption{\label{fig_n_lda}
Density of a trapped Fermi gas in the local density approximation
at a temperature $T/T_F=0.25$, slightly above the critical 
temperature. The thick line is the result based on the equation 
of state from Nascimbene et al., the dashed line is the result
for a free gas, and the dotted line shows the result using the 
equation of state in the high temperature limit. }   
\end{figure}

 In this appendix we describe a parameterization of the equation 
of state in the normal phase. The equation of state has been 
studied experimentally \cite{Luo:2008,Chevy:2009,Horikoshi:2010}, 
using quantum Monte Carlo simulations 
\cite{Lee:2005it,Burovski:2006,Bulgac:2008zz},
and many-body theory \cite{Chen:2005,Hausmann:2007}.
Here we follow the recent work of Nascimbene et al.~\cite{Chevy:2009} 
and write 
\be 
\label{eos_h}
P(\mu,T) = P_1(\mu,T) h(\zeta)\, ,
\ee
where $P_1(\mu,T)$ is the ideal gas equation of state of a single 
species non-relativistic Fermi gas
\be
\label{P_1}
 P_1(\mu,T) = -T\lambda_{\it dB}^{-3} {\it Li}_{5/2}(-\zeta^{-1})\, ,
\ee
and $\lambda_{\it dB}=[(2\pi)/(mT)]^{1/2}$ is the de Broglie
wave length. Here, ${\it Li}_{\alpha}(x)$ is the Polylogarithm 
function, and $\zeta=\exp(-\mu/T)$ is the fugacity. We parameterize 
$h(\zeta)$ as
\be 
\label{h_zeta}
\frac{h(\zeta)}{2} = \frac{\zeta^2+c_1\zeta+c_2}{\zeta^2+c_3\zeta+c_4}\, , 
\ee
and determine the parameters $c_{i}$ from a fit to the data 
of Nascimbene et al.~\cite{Chevy:2009}. This parameterization 
is motivated by the fact that the data for $\zeta>1$ is very well 
described by the Virial expansion $h(\zeta)/2=1+b_2/\zeta+b_3/\zeta^2+
O(1/\zeta^3)$. At unitarity $b_2=1/\sqrt{2}$ and $b_3=\frac{1}{8}-0.355
=0.23$. The value of $h(\zeta)$ at zero fugacity is related to the 
Bertsch parameter $\xi=\mu/E_F$. Using $\xi\simeq 0.4$ we have 
$h(0)/2=\xi^{-3/2}\simeq 3.8$. A fit for fugacities in the range 
$\zeta\in [0.03,5]$
gives 
\be 
c_1= 1.3543,\;\;
c_2=-0.0174,\;\;
c_3= 0.5724,\;\;
c_4=-0.0084.
\ee
We compare the data to our fit and the Virial expansion in Fig.~\ref{fig_eos}.
From the pressure we can determine other thermodynamic quantities. The 
density and entropy density are given by
\be 
n(\mu,T) = \lambda_{\it dB}^{-3} g(\zeta)\, , \hspace{0.5cm}
s(\mu,T) = \lambda_{\it dB}^{-3} k(\zeta)\, ,
\ee
with 
\bea
g(\zeta) &=& -{\it Li}_{3/2}(-\zeta^{-1})h(\zeta) 
           +\zeta{\it Li}_{5/2}(-\zeta^{-1})h'(\zeta)\, , \\
k(\zeta) &=& -\left(\log(\zeta){\it Li}_{3/2}(-\zeta^{-1}) 
                  +\frac{5}{2}{\it Li}_{5/2}(-\zeta^{-1}) \right)h(\zeta)
 \nonumber \\ 
    & & \mbox{}
                  +\log(\zeta){\it Li}_{5/2}(-\zeta^{-1})h'(\zeta)\, .
\eea
In a trapped system we use the local density approximation $n(x)=n(\mu(x),
T)$ with $\mu(x)=\mu-V(x)$ where $V(x)$ is the trapping potential. This 
determines the density profile if the temperature and the chemical 
potential (or the fugacity) at the center of the trap are given. In 
practice we usually specify the temperature and the total number 
of particles. The particle number defines a temperature scale $T_F
=(3N)^{1/3}\bar{\omega}$, where $\bar{\omega}=(\omega_x\omega_y
\omega_z)^{1/3}$ is the geometric mean of the trap frequencies. Given
$T/T_F$ the fugacity $\zeta_0$ at the center of the trap is determined
by the condition
\be 
 \frac{3}{(2\pi)^{3/2}}\left(\frac{T}{T_F}\right)^3 
\int d^3x\, g\left( \zeta_0 \exp\left( \frac{x^2}{2}\right)\right) 
 \equiv 1 \, . 
\ee
This equation has to be solved numerically. In the high temperature 
limit $\zeta_0= 6(T/T_F)^3$. In Fig.~\ref{fig_n_lda} we show the 
density profile at $T/T_F=0.25$. We show the exact density, the 
density of a free gas, and the high temperature (Gaussian) 
approximation. The effects of quantum degeneracy decrease the 
central density, whereas interactions increase the density. The 
two effects partially cancel and the exact density is about 50\% 
larger than the Gaussian approximation. 

 Once the initial density and pressure have been determined the 
equations of fluid dynamics fix the evolution of $P$ and $n$. The 
equation of state is needed in order to compute other thermodynamic
quantities like the temperature and the chemical potential \footnote{If
thermal conductivity is included then the temperature has to be 
determined at each step in the hydrodynamic evolution in order 
to compute the energy current.}. The fugacity can be computed 
from 
\be 
\label{fug}
 \frac{2}{(2\pi)^{3/2}} \left( \frac{mP}{n^{5/3}}\right)^{3/2}
 = 2\, \frac{f(\zeta)^{3/2}}{[-\zeta f'(\zeta)]^{5/2}}
 \equiv F(\zeta)\, , 
\ee
where $f(\zeta)=-{\it Li}_{5/2}(-\zeta^{-1})h(\zeta)$.
Equ.~(\ref{fug}) implies that 
\be 
\zeta = F^{-1}\left( \frac{2}{(2\pi)^{3/2}} 
      \frac{m^{3/2}P^{3/2}}{n^{5/2}}\right)\,. 
\ee
In general, $F^{-1}(y)$ has to computed numerically. In the 
high temperature limit $F^{-1}(y)\simeq y$. Once the fugacity 
is known the temperature can be computed from 
\be 
 T =- \frac{\zeta f'(\zeta)}{f(\zeta)} \frac{P}{n}\, . 
\ee
In the high temperature limit $f(\zeta)\simeq 2/\zeta$ which
implies $\zeta f'(\zeta)/f(\zeta)\simeq -1$ and $T=P/n$. 


\end{document}